\newif\ifarxiv
\arxivtrue   % arXiv preprint build (mode 3). \arxivfalse returns to the CJ/WWW tracks.
\newif\ifcj
\cjtrue    % CJ track = KDD Customer-Journey workshop (\cjfalse = arXiv/WWW)
\newif\ifcamera
\cameratrue % CJ track only: \cameratrue = accepted camera-ready, \camerafalse = v1 submission
\newif\ifshowethics
\ifcj\showethicstrue\else\ifarxiv\else\showethicstrue\fi\fi
\newif\ifvtwo
\ifcj\ifcamera\vtwotrue\fi\fi
\ifarxiv\vtwotrue\fi
\ifarxiv
  \documentclass[sigconf, nonacm]{acmart}
\else\ifcj\ifcamera
  \documentclass[sigconf]{acmart}         % (1) CJ@KDD '26 camera-ready: NO line numbers
\else
  \documentclass[sigconf, review]{acmart} % (2) CJ@KDD '26 v1 submission: line numbers
\fi\else
  \documentclass[sigconf, review]{acmart} % (4) WWW industry-track review submission
\fi\fi
\AtBeginDocument{}
\usepackage{tikz}
\usetikzlibrary{arrows.meta,positioning,fit,backgrounds}
\usepackage{microtype}
\usepackage{enumitem}
\ifcj\ifcamera
  \setcopyright{none}              % TODO eRights: -> e.g. \setcopyright{acmlicensed}
\else
  \setcopyright{none}
\fi\else
  \setcopyright{none}
\fi
\renewcommand\footnotetextcopyrightpermission[1]{}
\ifcj
  \ifcamera
    \acmConference[CJ@KDD~'26]{5th Workshop on End-to-End Customer Journey Optimization, KDD~'26}{August~9, 2026}{Jeju, Republic of Korea}
  \else
    \acmConference[Submitted to CJ@KDD~'26]{5th Workshop on End-to-End Customer Journey Optimization, KDD~'26}{August~2026}{Jeju, Korea}
  \fi
  \acmYear{2026}
\else
  \ifarxiv
    \acmConference[Preprint]{Preprint --- not peer reviewed}{}{}
  \else
    \acmConference[WWW '27]{The ACM Web Conference 2027}{2027}{}
  \fi
  \acmYear{2027}
\fi

\begin{document}

\title{From Prompt to Purchase: How AI Brand Recommendations Move Consumers on the
Open Web}
% Acceptance note: shown on the CJ camera-ready and the arXiv preprint, NOT on the
% frozen v1 submission (single-blind) or the WWW submission. The two wordings are
% mutually exclusive; \ifarxiv is tested first because the arXiv build (mode 3) is
% reached with \cjtrue still set, which would otherwise emit both notes.
\ifarxiv
  \titlenote{A version of this paper was accepted as a Contributed Talk at the 5th
  Workshop on End-to-End Customer Journey Optimization (CJ@KDD~'26).}
\else\ifcj\ifcamera
  \titlenote{Accepted as a Contributed Talk at the 5th Workshop on End-to-End
  Customer Journey Optimization (CJ@KDD~'26), Jeju, Republic of Korea.}
\fi\fi\fi

\author{Michael Iannelli}
\orcid{0000-0002-5967-2026}
\affiliation{%
  \institution{Scrunch AI}
  \city{New York}
  \country{USA}}
\email{michael@scrunchai.com}

\author{Alan Ai}
\orcid{0009-0004-8467-1633}
\affiliation{%
  \institution{Scrunch AI}
  \city{New York}
  \country{USA}}
\email{alan.ai@scrunchai.com}
\renewcommand{\shortauthors}{Iannelli and Ai}

\begin{abstract}
When a conversational assistant recommends a brand to a user with no recent
observed engagement, that user's same-name Google search rises $+4.3$ percentage
points (pp) [$3.1$, $5.5$], visits to the brand's own site $+2.4$~pp [$1.4$, $3.5$], and
brand-specific retailer-page visits $+1.0$~pp [$0.3$, $1.7$] over matched backward
placebos. Recovering that estimate is the work. The mention creates a brand
exposure no web log attributes to the assistant, and the naive all-mention funnel
that seems to measure it is confounded: many mentions are incidental references to
brands the user already uses (``your Netflix download''), whose downstream visits
are that existing customer's own behavior and surface as a brand-specific
pre-trend. We measure off-platform response on a panel that joins opt-in clickstream
to the same users' ChatGPT, Claude, and Gemini conversations, and isolate the effect
with a pre-trend event study, a stance classifier, non-customer conditioning, and a
within-response same-category control: incidental name-drops then move behavior far
less ($+1.8/+1.1/+0.3$), and the named brand moves far more than unnamed
same-category brands in the same response. The downstream path is mostly
search-mediated and reaches both own sites and retailer pages, with a destination
mix that tracks baseline brand-directed behavior rather than redirecting toward
either. The design is observational and we do not observe transactions, so retail is
purchase-adjacent. Standard referrer-based and last-click measurement miss this upstream
exposure: assistants move observably-unengaged users into open-web brand navigation
along a path attributed elsewhere\ifcj{} --- an acquisition touchpoint at the head of the
customer journey that journey models and last-click attribution do not see\fi.
\end{abstract}

\ifcj
\else
\fi

\ifarxiv\else
\fi

\maketitle

\section{Introduction}

Ask ChatGPT ``what running watch should I buy?'' and it answers ``Garmin, Coros,
or Polar.'' That sentence is now a routine on-ramp to the web: the user reads
three brands they did not type, and some of them go looking. The assistant has
become a step in the path to commerce that no web log records. Last-click
attribution sees the eventual Google search or retailer visit and credits
organic channels; the antecedent, the brand the model named, is invisible. As
conversational assistants---the text-based LLM chat interfaces ChatGPT, Claude, and
Gemini, as distinct from voice assistants and from search-embedded AI like Google's
AI Overviews, which we exclude (\S\ref{sec:data})---absorb a growing share of
information-seeking~\cite{padilla2025,gholami2026,chapekis2025}, that blind spot
widens: the brand the model named pulls the user toward it along a path no log
attributes to the assistant.

\begin{figure*}[t]
\centering
\begin{tikzpicture}[
  node distance=4mm and 8mm, font=\footnotesize,
  box/.style={draw, rounded corners=2pt, align=center, inner sep=4pt, minimum height=9mm},
  prompt/.style={box, fill=blue!8, draw=blue!50, text width=28mm},
  llm/.style={box, fill=orange!10, draw=orange!60, align=left, text width=40mm},
  stage/.style={box, fill=green!8, draw=green!50!black, align=left, text width=36mm},
  ->, >={Stealth[scale=0.9]}, thick]
\node[prompt] (p) {User prompt\\\textit{``best running watch?''}};
\node[llm, right=of p] (r) {Assistant response:\\\textit{``Consider \textbf{Garmin}, \textbf{Coros}, or \textbf{Polar}.''}};
\node[stage, right=of r] (rec) {\textbf{Recall} $+4.3$pp\\search ``garmin''};
\node[stage, right=of rec, yshift=9mm] (disc) {\textbf{Discovery} $+2.4$pp\\visit \underline{garmin.com}};
\node[stage, right=of rec, yshift=-9mm] (ret) {\textbf{Retail} $+1.0$pp\\garmin page at a retailer};
\draw (p) -- (r); \draw (r) -- (rec);
\draw (rec.east) -- ++(2.5mm,0) |- (disc.west);
\draw (rec.east) -- ++(2.5mm,0) |- (ret.west);
\end{tikzpicture}
\Description{A user prompt asking for the best running watch leads to an assistant response naming Garmin, Coros, and Polar. When the assistant recommends a brand to a non-customer, the response is followed by a same-name search (recall, +4.3 percentage points), the entry point, from which two parallel destinations branch off: a visit to the brand's own site (discovery, +2.4) and a brand-specific retailer page visit (retail, +1.0). The stages co-occur strongly but the own-site and retailer visits are not ordered. None is logged as caused by the assistant.}
\caption{The clean acquisition effect. A same-name search is the entry point;
own-site and retailer visits are \emph{parallel} destinations that co-occur without a fixed order
(\S\ref{sec:structure}). Lifts are recommendation $\times$ non-customer
(\S\ref{sec:clean}); standard analytics attribute almost none to the assistant.}
\label{fig:flow}
\end{figure*}
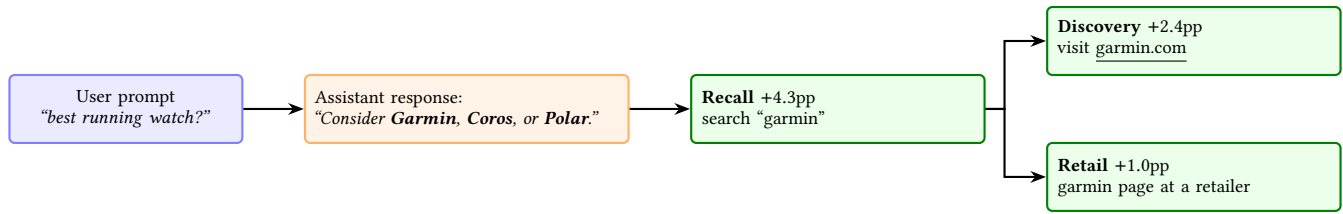

We measure that path on a deployed measurement panel that joins opt-in pageview
clickstream to the same panelists' conversations under strict aggregation
constraints, and which has run in production at Scrunch~AI since early 2026: a same-name
search, a visit to the brand's own site, a brand-specific retailer-page visit
(Figure~\ref{fig:flow}). The headline is an acquisition-like
effect.\footnote{Three terms recur, all defined behaviorally. A
\emph{non-customer} (observably-unengaged user) is one with no recent observed
search, own-site, or retailer engagement with the brand, not someone proven never
to have used it. \emph{Retail} is a purchase-adjacent retailer product-page visit,
not an observed transaction. \emph{Acquisition-like} denotes movement among
observably-unengaged users, not a measured purchase or sign-up.} When the
assistant \emph{recommends} a brand to a user with no recent observed engagement
with it, the user searches it ($+4.3$~pp), visits its site ($+2.4$~pp), and
reaches it at a retailer ($+1.0$~pp), all over a within-user backward placebo. The
design is observational and we do not observe transactions; the retailer-page
visit is purchase-adjacent. Standard web analytics attributes almost none of these to the
assistant.

This clean number has to be separated from a confounded one, and that
separation is the contribution. Pooling every mention gives an attractive but
confounded funnel ($+2.08/+2.37/+0.52$~pp); it is not the estimand we want. The
pooled lift averages genuine recommendations with incidental
references to brands the user already uses (``you can use Samsung Pay,'' ``your
Netflix download''), and for those the downstream visits are the user's
\emph{existing-customer} behavior, already underway, so the pooled discovery and
retail lifts are biased upward as estimates of what a mention causes. An event study makes the
bias visible. The named brand's own-site activity is already rising in the days
\emph{before} the response, while same-category brands the response did not name
stay flat. Conditioning on non-customers and on genuine recommendations removes
this pre-trend and recovers the acquisition effect above; incidental name-drops
then move behavior two to three times less, and the measurable effect concentrates
among users with no recent observed engagement (among already-engaged users the
mention coincides with an active episode without clearly accelerating it). A reverse path runs the
other way, with existing brand-directed demand flowing
\emph{into} the conversation, and the same conditioning removes it so the
forward prompt-to-purchase contrast becomes cleaner.

\paragraph{Contributions.}
The claim in one line: conversational assistants create unlogged brand exposures that
move observably-unengaged users into open-web brand navigation.
(1) A method for measuring the off-platform behavioral response to an LLM-generated
brand exposure under observational constraints, and the acquisition-like effect it
recovers: on a joined clickstream--conversation panel, an AI recommendation to a
non-customer is followed by more same-name search, own-site discovery, and
purchase-adjacent retail, almost none of it visible to last-click attribution
(\S\ref{sec:clean}). (2) A decomposition that separates this effect
from two confounds, incidental name-drops and existing-customer behavior (a
brand-specific pre-trend), using a pre-trend event-study, non-customer
conditioning, a stance classifier, and a within-response same-category control
(\S\ref{sec:catctrl}--\S\ref{sec:naive}). (3) Practical measurement corrections and
scope conditions: path-aware retail matching and platform-brand separation recover a
retail effect that host-only matching reads as near-null, while heterogeneity analyses
show category-specific destinations but no reliable familiarity or mention-position
gradient (\S\ref{sec:measure}--\S\ref{sec:second}).
\ifcj
The framing is for customer-journey optimization: the assistant is an unlogged \emph{acquisition}
touchpoint at the awareness-to-acquisition boundary --- it moves people who were not already in
the funnel --- so journey models and uplift measurement that omit it misattribute acquisition to
downstream organic channels. In customer-lifecycle terms the effect sits where broad-audience
brand awareness feeds first-time acquisition; the recommendation\,$\times$\,non-customer contrast
is an uplift estimand (incremental lift on the observably-unengaged), and the existing-customer
analysis separates acquisition-like reach from mature-user activity. For journey practitioners the implication is
concrete: this is a new, currently-unattributed acquisition channel whose incremental lift on
non-customers --- not the confounded all-mention funnel --- is the quantity that channel-investment
and targeting decisions need, and that last-click attribution never surfaces.
\else
The framing is for the web-measurement community: the assistant is an unlogged hop in
the path to commerce, and attribution that ignores it misallocates credit.
\fi

\section{Related Work}

\paragraph{AI-mediated web behavior.} A growing body of panel work documents how
conversational assistants reshape web traffic in aggregate. LLM adoption shifts
the volume and concentration of where users go~\cite{padilla2025,gholami2026},
and SERP-level AI summaries depress organic click-through~\cite{chapekis2025}.
Closest to a commercial outcome, Kaiser and Schulze~\cite{kaiser2026referrals} measure observable
ChatGPT \emph{referral} traffic and transactions reaching e-commerce sites
through in-answer links, and find the channel emerging but still small. Our
setting is complementary: we measure brand \emph{mentions} that usually generate
no observable assistant referral, and show they move users through ordinary
search and brand-navigation paths that last-click attribution assigns elsewhere.
These are aggregate, short-run channel
flows: they establish that assistants redirect traffic, but not where a
\emph{specific} mention sends a \emph{specific} user, nor whether it moves people
who were not already headed there. We fill that gap at the level of the
individual exposure.

\paragraph{Web attribution and incrementality.} The central problem of digital
attribution is that the click which converts is rarely the touch that caused the
conversion. Last-click analytics credits the final channel and is blind to
upper-funnel exposure; multi-touch attribution spreads credit across the
path~\cite{berman2018lasttouch}, and the incrementality literature shows, in large
field experiments, that observational credit and true causal lift diverge by an
order of magnitude, with paid search and display often capturing demand that
would have converted anyway~\cite{lewis2014,blake2015,gordon2019comparison}. That
divergence motivated ghost-ad and intention-to-treat designs that measure the
counterfactual directly~\cite{johnson2017ghost}. Television and other advertising
exposures have long been studied through downstream search, purchase, or response
footprints~\cite{joo2014tvsearch,sahni2015}. The AI mention is a new
member of this family of hard-to-log upper-funnel touches, with a twist: it is
not even recorded as an exposure. Our central methodological move, separating a
mention's effect from the pre-existing demand it co-occurs with, is the
incrementality question in a setting where the treatment is unlogged.

\paragraph{The consumer search journey and omnichannel retail.} Our three stages,
search, own-site, and retailer, trace the recall, consideration, and purchase
spine of the decision-journey
literature~\cite{court2009journey,lemon2016journey}, and the multi-surface
routing we observe (a brand site for some categories, a third-party retailer for
others) is the omnichannel reality a single funnel number
hides~\cite{verhoef2015omnichannel}. Online product recommendations move consumer
choice when presented as recommendations~\cite{senecal2004influence}; the
assistant's mentions are different in kind, incidental and in prose, and we
separate the ones that function as recommendations from those that do not.

\paragraph{Assistants as recommenders.} An assistant that names entities is, in
one reading, a recommender without a recommendation UI. A line of work studies
LLMs as rankers~\cite{hou2024,bao2023tallrec,geng2022p5} and their popularity and
position biases~\cite{lichtenberg2024}, and the
conversational-recommendation literature studies preference elicitation in
dialogue~\cite{christakopoulou2016,jannach2021survey}. Our novelty is downstream
and off-platform. Two findings speak back to that literature: the long-tail value
case~\cite{anderson2006longtail,brynjolfsson2003variety,fleder2009diversity}
predicts larger effects for less-familiar items, which we test and do not find;
and whereas on-platform ranked lists show strong position
bias~\cite{lichtenberg2024}, the off-platform behavioral effect is flat across
mention order (\S\ref{sec:second}).

\section{Data and Methods}
\label{sec:data}

\paragraph{Panel and disclosure.} An opt-in clickstream panel covers two
English-speaking markets with pageview-level capture, joined per user at
timestamp granularity to the same panelists' conversations with ChatGPT, Claude,
and Gemini. This measurement layer has run in production at Scrunch~AI since early
2026, supporting report-form analyses of AI-mediated consumer journeys; the paper
reports a disclosure-safe aggregate of the same deployed pipeline (\S\ref{sec:deploy}).
We do not report aggregate panel size, per-analysis user counts, or
per-cell sample sizes; these are commercially sensitive. Effect sizes, confidence
intervals, rates, and ratios are reported throughout; every headline estimate is a
user-clustered bootstrap over the full eligible panel rather than a sampled subset.
Every headline cell clears the platform's minimum-disclosure threshold on distinct
users, distinct brands, and user--response--brand rows by a wide margin, so no
reported interval rests on a thin cell. Search-embedded surfaces
(Google AI Overview, AI Mode) are excluded: their ``prompt'' is a search the user
just typed, so a same-name search outcome would be mechanical.

\paragraph{Unit, outcomes, identification.} The unit is a (user, response, brand)
row for a curated cross-category brand lexicon (high recognition, low homograph
ambiguity) spanning consumer categories; three additional single-category verticals
are introduced only for the generalization test in \S\ref{sec:second}.
Three outcomes are measured over the seven days after a response: \emph{recall}
(brand recall, operationalized as a Google search whose query contains the brand---not
the information-retrieval metric), \emph{discovery} (a visit to
the brand's own site), and \emph{retail} (a visit to a retailer page whose URL
path identifies the brand). For each row we difference the outcome over
$(T, T{+}w)$ against the same user's rate over same-width windows at $T{-}14$,
$T{-}21$, $T{-}28$ days. Matched widths are necessary: differencing a 24h
outcome against a 7-day placebo biases the contrast negative by construction, so
every contrast uses windows of equal width. Formally, for an eligible exposure set
$\mathcal{E}$ (e.g.\ recommendation $\times$ non-customer), the lift over window $w$ is
\begin{multline*}
\hat{\Delta}_{w} = \frac{1}{|\mathcal{E}|}\sum_{(i,r,b)\in\mathcal{E}}\Big[\,Y_{i,r,b}(T_r,\,T_r{+}w) \\
{}- \tfrac{1}{3}\sum_{k\in\{14,21,28\}} Y_{i,r,b}(T_r{-}k,\,T_r{-}k{+}w)\,\Big],
\end{multline*}
where $i,r,b$ index user, response, and brand, $T_r$ is the response time, and
$Y_{i,r,b}(\cdot,\cdot)=1$ if any qualifying event occurs in the window. A novelty filter (no same-brand
\emph{search} in the prior week) conditions every stage. Intervals are
user-clustered bootstraps. This is observational; ``lift'' is an event-study
contrast under observational data, with no randomized exposure. Table~\ref{tab:windows} maps the time windows.

\begin{table}[t]
\caption{Time windows, all anchored on the response at $T$.}
\label{tab:windows}
\small
\begin{tabular}{@{}l p{0.56\columnwidth}@{}}
\toprule
Window & Purpose \\
\midrule
Prior week & novelty filter (no prior same-brand search) \\
Prior week & non-customer eligibility (no observed search, own-site, or retail) \\
$T{-}14/{-}21/{-}28$ & matched backward-placebo windows \\
$T$ to $T{+}7$ & headline post-response outcome window \\
$29$--$56$ d before $T$ & existing-customer definition (non-overlapping, \S\ref{sec:exist}) \\
\bottomrule
\end{tabular}
\end{table}

\paragraph{Cleaning the contrast.} The novelty filter blocks prior search but not
prior site or retail activity, so it does not by itself remove existing-customer
behavior; the controls that do are introduced with the results they support
(\S\ref{sec:clean}--\S\ref{sec:naive}). We classify each mention's \emph{stance}
toward the named brand (recommend / neutral / caution) with a small language
model, and define a \emph{non-customer} of a brand as a user with no recent
\emph{observed} engagement: no prior search, own-site, or retail pageview in the
prior window. This is a behavioral definition, not a commercial one. Such a user
may still own the product, buy offline, or engage through channels the panel does
not capture. We call such a user a \emph{non-customer} for readability, but the
estimand is precisely a user with no recent \emph{observed} brand engagement (an
\emph{observably-unengaged} user), and we read the effect as \emph{acquisition-like}
(\S\ref{sec:limits}). Our cleanest observational contrast is a recommendation to
such a user.

\paragraph{Measuring retail.} Retail counts only when the retailer URL identifies
the brand, and the brand lives in the URL \emph{path}
(\texttt{.../Garmin-.../dp/}), not the host. Matching the host alone captures a
small minority of product-page visits and reads as a near-null, so we match the
path. Canonical retailer URLs (e.g.\ an Amazon \texttt{/dp/} product page) carry no
brand string, so path-aware brand-in-URL visits are the primary retail measure,
with the search-anchored union an upper bound only.

\section{Results}
\label{sec:results}

Every threat to this design is a way the pooled lift could arise without the
assistant moving the user. Table~\ref{tab:threats} lists each, the control that
addresses it, and the result that does so; the one threat we cannot fully close on
observational data, a brand-specific within-session intent shock, is stated as
such (\S\ref{sec:limits}). The subsections below report the clean effect first,
then the controls and the decomposition that produce it.

\begin{table*}[t]
\caption{Identification and measurement threats and the control that addresses each; the evidence
column gives the diagnostic or result that addresses it. The one residual threat, a brand-specific
within-session intent shock, is not fully identifiable on observational data
(\S\ref{sec:limits}).}
\label{tab:threats}
\small
\begin{tabular}{p{0.20\linewidth}p{0.30\linewidth}p{0.42\linewidth}}
\toprule
\textbf{Threat} & \textbf{Control} & \textbf{Evidence / status} \\
\midrule
Existing-customer activity already underway & Pre-trend event-study; non-customer conditioning & The named-brand pre-trend ($+1.68$~pp discovery before the response) vanishes in the clean cell (\S\ref{sec:naive}) \\
Incidental name-drops are not recommendations & Stance classifier (recommend / neutral / caution) & Recommendations move two to three times more than neutral name-drops (\S\ref{sec:clean}) \\
Category intent the user brought to the session & Within-response same-category control; no-prior-category stratum & Unnamed same-category brands stay near flat; the effect holds for users not browsing the category (\S\ref{sec:catctrl}) \\
General activity burst on assistant-session days & Within-response same-category control (shared session) & Both the named and unnamed brands share the session, yet only the named brand moves (\S\ref{sec:catctrl}) \\
Retail URL measurement error & Path-aware brand-in-URL matching & Host-only retail reads near-null; path-aware matching recovers $+0.52$~pp (\S\ref{sec:measure}) \\
Brand-specific within-session intent shock & No-prior-category stratum bounds it; otherwise unidentified & The effect survives the lowest-intent stratum, but a within-session shock is not closable without a randomized design (\S\ref{sec:limits}) \\
\bottomrule
\end{tabular}
\end{table*}

\subsection{An acquisition-like lift among observably-unengaged users}
\label{sec:clean}

\begin{figure}[t]
\centering
\includegraphics[width=0.86\linewidth]{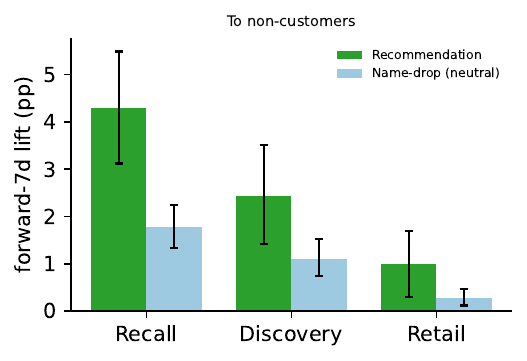}
\Description{Grouped bar chart of the forward seven-day lift for recall,
discovery, and retail, comparing recommendations to neutral name-drops, both for
non-customers. Recommendations are about two to three times larger at every
stage, with confidence intervals above zero.}
\caption{The acquisition effect and its stance dose, among non-customers. A
recommendation moves all three stages two to three times more than an incidental
name-drop. 95\% user-clustered bootstrap CIs.}
\label{fig:acqbars}
\end{figure}

When the assistant \emph{recommends} a brand to a non-customer, all three downstream
behaviors rise (Table~\ref{tab:clean}, Figure~\ref{fig:acqbars}). Recall rises
$+4.3$~pp [$3.1$, $5.5$], own-site discovery $+2.4$~pp [$1.4$, $3.5$], and
brand-specific retail $+1.0$~pp [$0.3$, $1.7$]. This cell's pre-window is zero
by the non-customer construction, so we do not read its flat pre-trend as
behavioral evidence; identification rests on the matched backward placebo and the
same-response control (\S\ref{sec:catctrl}), and the step at the response is clean
of either (Figure~\ref{fig:eventstudy}, green). The
effect scales with stance. An incidental name-drop (a neutral mention) to the
same kind of non-customer moves behavior far less ($+1.8/+1.1/+0.3$): being named
at all does something, but a recommendation does two to three times more. The
pattern is consistent with the assistant acting as an upper-funnel touch that
introduces or re-surfaces a brand to someone with no recent observed engagement.

\begin{table}[t]
\caption{Observed-navigation rates and lift by stance among users with no recent
observed brand engagement. Baseline is the matched within-user backward placebo and
treated is the 7-day post-response window (both rates, \%); lift is their difference
(pp). A recommendation moves the funnel two to three times more than an incidental
name-drop. Each stance arm has its own matched backward-placebo baseline (they are
different exposures), and recommended brands sit on modestly higher baselines.
Existing-customer acceleration is a separate
estimand (\S\ref{sec:exist}). Recommend lift CIs: recall [$3.1,5.5$], discovery
[$1.4,3.5$], retail [$0.3,1.7$].}
\label{tab:clean}
\small
\begin{tabular}{lccc}
\toprule
Stage & Baseline & Treated & Lift (pp) \\
\midrule
\multicolumn{4}{@{}l}{\emph{Recommendation, no recent engagement}} \\
Recall    & $2.9$ & $7.2$ & $+4.3$ \\
Discovery & $3.1$ & $5.5$ & $+2.4$ \\
Retail    & $0.8$ & $1.8$ & $+1.0$ \\
\midrule
\multicolumn{4}{@{}l}{\emph{Name-drop (neutral), no recent engagement}} \\
Recall    & $2.5$ & $4.3$ & $+1.8$ \\
Discovery & $2.7$ & $3.8$ & $+1.1$ \\
Retail    & $0.4$ & $0.7$ & $+0.3$ \\
\bottomrule
\end{tabular}
\end{table}

\subsection{The lift is specific to the named brand}
\label{sec:catctrl}

A session in which the assistant names a brand is also a session of higher
brand-directed intent and activity, so the lift could reflect the user's own agenda
rather than the mention. Four controls narrow this alternative, and the effect
survives each.

\emph{The named brand, in the same response.} The sharpest control holds the
response fixed. Within the responses that contain a recommended non-customer
brand, that brand draws recall $+4.4$, discovery $+2.7$, retail $+0.8$, while the
\emph{unnamed} same-category brands in the same responses (also with no recent
observed engagement) draw $+0.6$, $+0.3$, $+0.3$: an order of magnitude smaller at
recall and discovery. The pattern is specific to the named brand, beyond the
category or session, and not regression among low-baseline users. (Retail is
noisier, where the named-vs-control gap is directional with overlapping
intervals; a broader same-category control pool gives $+0.3/+0.0/+0.1$.) Because
this control needs lexicon substitutes for the brand, it runs on the
category-comparable subset, whose treated estimates ($+4.4/+2.7/+0.8$) track the
all-brand headline.

\emph{The category, across all mentions.} The same control at the level of all
mentions tells the same story (Table~\ref{tab:catctrl}): the named brand shows
the funnel while same-category brands the response did not name barely move
(recall $+2.06$ vs $+0.20$; discovery $+2.39$ vs $+0.09$). Because the named and unnamed brands share
the response and session, this addresses two confounds at once: a time-varying
\emph{category} intent (the user wanted a phone this week) and a general
\emph{activity} surge (the days a user converses with an assistant are
higher-activity days). Either would lift the unnamed brands too, and neither
does. Retail is the one stage with a small non-zero control ($+0.16$ vs $+0.51$),
consistent with the activity confound that inflates retail at large, leaving a
brand-specific retail effect of about $+0.35$~pp net of it.

\emph{No prior observed category activity.} A brand-specific intent the user brought
to the session by browsing the category beforehand could remain. Among
category-comparable responses, restricting to users with no observed prior
same-category \emph{lexicon-brand} engagement in the prior week (no own-site,
retailer, or search activity for any mapped same-category brand) retains $92\%$ of
the cell and leaves the effect intact (recall $+4.3$ [$3.0$, $5.5$], discovery $+2.9$
[$1.8$, $3.9$], retail $+0.8$ [$0.1$, $1.5$]). This narrows, but does not eliminate,
prior intent: it does not capture generic category searches, non-lexicon brands, or
unobserved or offline research.

\emph{A stricter window and the assistant-volunteered cut.} The effect
strengthens under a 28-day non-customer window ($+5.4/+4.5/+1.0$) and holds for
\emph{unprompted} recommendations ($+3.4/+2.2/+0.7$), where the assistant
volunteered the brand and the user did not name it, easing the concern that the
user drove the exposure.

\begin{table}[t]
\caption{Within-response same-category control (all mentions): named brand vs the
same-category brands the response did \emph{not} name. Within-user
backward-placebo lift (pp, 7d). Recall/discovery controls are $\approx 0$; retail
carries a small bleed.}
\label{tab:catctrl}
\small
\begin{tabular}{lcc}
\toprule
Stage & Named brand & Unnamed same-cat. \\
\midrule
Recall    & $+2.06$ [$1.61$,$2.57$] & $+0.20$ [$-0.00$,$0.40$] \\
Discovery & $+2.39$ [$1.75$,$2.99$] & $+0.09$ [$-0.42$,$0.44$] \\
Retail    & $+0.51$ [$0.25$,$0.75$] & $+0.16$ [$0.04$,$0.28$] \\
\bottomrule
\end{tabular}
\end{table}

\subsection{Acquisition-like reach: the effect concentrates among non-customers}
\label{sec:exist}

If the assistant acted as an in-store salesperson, closing a sale the user was
already pursuing, existing customers would jump after a mention beyond their own
rising baseline. The picture is more subtle. Defining existing customers by site or
retail engagement $29$--$56$ days before the response (a window chosen so it does
\emph{not} overlap the backward placebos, unlike an $8$--$28$d window that overlaps
them and mechanically depresses the contrast), they are already in an active brand
episode around the response: their 3-day post-response discovery runs $+4.7$~pp
[$+1.8$, $+7.7$] and retail $+0.3$~pp [$-0.6$, $+1.3$] above the matched placebo. But
the episode is already underway. Comparing the 3 days after the mention to the 3
days before, the added increment is small (discovery $+0.4$~pp, retail $\approx 0$):
the mention coincides with the episode more than it accelerates it. The
recommend-to-existing subset hints at a larger post-mention increment, but that cell
is small and its interval wide, so we do not rest a claim on it. We define existing engagement here by site or retail activity rather than prior
search, because the estimand is acceleration of downstream navigation, not recall; and
we use a 3-day window to compare immediate post-mention movement against the
immediately preceding episode, while the headline acquisition analysis stays at 7
days. The post-minus-pre increment is descriptive: we do not attach a separate
bootstrap interval to it.

The assistant's cleanly measurable downstream effect is therefore concentrated among
users with no recent observed engagement (\S\ref{sec:clean}); among already-engaged
users the mention sits within an active episode, and we cannot cleanly resolve
whether it adds a modest acceleration on top. The measurable role is closer to
upper-funnel reach than to closing an already-active episode: what we can measure is
reach to observably-unengaged users (\S\ref{sec:deploy}). Two caveats remain. ``Existing
customer'' is defined by observed engagement, which undercounts affinity that never
appears as a click. And the reverse path, existing demand flowing \emph{into} the
conversation, is the pre-trend we remove (\S\ref{sec:naive}), not a second causal
channel.

\subsection{Why the pooled funnel is confounded}
\label{sec:naive}
\label{sec:pretrend}

\begin{figure}[t]
\centering
\includegraphics[width=0.99\linewidth]{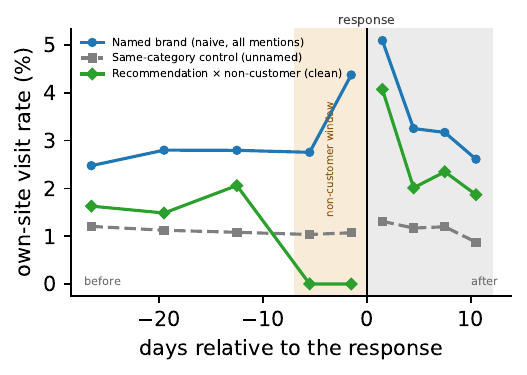}
\Description{Own-site visit rate by day relative to the response. The naive
all-mentions line rises in the days before the response, peaks just after, and
decays, a pre-trend. The same-category control line is flat throughout. The
recommendation-times-non-customer line is flat near zero before the response and
steps up sharply at it, with no pre-trend.}
\caption{Own-site event study around the response. Naive all-mention named brands
(blue) rise \emph{before} the response, a pre-existing-customer pre-trend;
same-category unnamed brands (grey) are flat; recommendation $\times$ non-customer
(green) steps up only at the response. The peach band is the non-customer
eligibility window, zero by construction; identification compares the post-response
level to the backward placebo at $T{-}14/{-}21/{-}28$, not to this window
(\S\ref{sec:naive}). Curves are raw daily own-site visit rates; the placebo
comparison is applied in the analysis, not to the plotted values. Bins are 3 days at
bin center.}
\label{fig:eventstudy}
\end{figure}

The central identification problem is that many AI brand mentions are not exposures
to new brands; they are references to brands already active in the user's journey.
Pooling all mentions gives an attractive but confounded funnel
(Table~\ref{tab:funnel}): recall $+2.08$~pp within a week (and $+1.57$ at 24h,
in the range of prior panel estimates), discovery $+2.37$~pp, retail $+0.52$~pp. For the
all-mention estimand the discovery and retail lifts overstate what a mention causes,
and an event study shows why. In the three days \emph{before} a mention, the
named brand's own-site activity already runs $+1.68$~pp above the backward
placebo, and its retail activity $+0.24$~pp, whereas the same-category
brands the response did not name are flat (discovery $-0.07$, retail $-0.01$);
Figure~\ref{fig:eventstudy} shows the full event-study. The pre-trend is
brand-specific, and it is present whether the user introduced the brand or the
assistant volunteered it. Many of these ``mentions''
are incidental references to brands the user already uses, such as a question
about a Samsung phone case, a Netflix download, or ``cashback at Nike,'' and the
surrounding visits are that existing customer's own behavior. Differencing the
post-mention rate against the pre-trend and netting out the same-category control
leaves a much smaller mention-attributable increment for discovery
($\approx{+}0.5$~pp) and retail ($\approx{+}0.1$~pp). Recall is unaffected: the
novelty filter already removes prior searchers and its pre-trend is negative, so
recall is the least confounded stage as measured, while the downstream stages are not until
the existing-customer episodes are removed, which is exactly what the
non-customer conditioning of \S\ref{sec:clean} does.

\begin{table}[t]
\caption{The naive pooled funnel (product brands; within-user backward-placebo
lift, pp, 95\% user-clustered bootstrap CI; 7d, matched placebos, novelty
filter). These pool genuine recommendations, incidental name-drops, and
existing-customer behavior; the clean decomposition is \S\ref{sec:clean}.}
\label{tab:funnel}
\small
\begin{tabular}{lcc}
\toprule
Stage & Lift (pp) & Treated rate \\
\midrule
Recall (24h)            & $+1.57$ [$1.26$, $1.87$] & $2.22\%$ \\
Recall (7d)             & $+2.08$ [$1.55$, $2.54$] & $5.32\%$ \\
Discovery (7d)          & $+2.37$ [$1.74$, $2.89$] & $7.20\%$ \\
Retail (7d, path-aware) & $+0.52$ [$0.30$, $0.77$] & $1.18\%$ \\
\bottomrule
\end{tabular}
\end{table}

\subsection{One episode, not a strict sequence}
\label{sec:structure}

We characterize the navigation structure on the pooled product-brand funnel, which
is large enough to estimate co-occurrence and ordering, as descriptive evidence for
the navigation pattern around AI brand mentions; the clean recommendation $\times$
non-customer cell shows the same three-stage lift but is not separately used for the
ordering estimates. Across that funnel the three
stages are one coherent episode---neither three independent behaviors nor a strict
deterministic chain. They are tightly coupled: within seven days they co-occur far above
chance (recall--discovery $5.2\times$ their independent rate, discovery--retail
$7.2\times$, recall--retail $7.8\times$), and retail is $14\times$ more likely when
the user also visited the brand's site. The marginal rates show the episode is not
nested, since discovery ($7.2\%$) exceeds recall ($5.3\%$): more users reach the
brand's site than search its name. (Recall counts only same-name Google queries, so discovery can exceed it
even when search is the dominant route into brand destinations: a user may arrive
from a broader query or a remembered name.) And within the episode the commercial stages do
not order. Among co-firing events the same-name search precedes the retailer visit
$72\%$ of the time, yet the own-site visit precedes the retailer visit only $39\%$ of
the time. The picture is a search-anchored episode with own-site and retailer visits
as \emph{parallel} destinations.

\subsection{Where the lift lands}
\label{sec:measure}

In the pooled product-brand funnel, decomposed by destination, the lift is larger at the brand's own site (discovery,
$+2.37$~pp) than at third-party retailers ($+0.52$~pp on the same basis). This is
not evidence that the assistant \emph{routes} demand to own sites. Comparing the
treated and placebo windows like-for-like, the destination mix does not shift: the
Amazon-versus-other split of retail does not move (pooled $+2$~pp, 95\% CI
$[-7,+11]$), and the own-site share of own-plus-retail activity, if anything, edges
\emph{down} (from $88\%$ to $85\%$, $[-4.7,-0.2]$)---a slight tilt toward retail,
the opposite of disintermediation. Comparing the increment's per-response
\emph{incidence} share against baseline pageview \emph{volume} can mislead: Amazon
is $92\%$ of brand-matched retail volume but only about half of per-response retail
incidence, so the two are not comparable. We therefore report the decomposition as
a descriptive fact and make no welfare or disintermediation claim --- the assistant
amplifies brand-directed navigation across stages without redirecting its
destination mix. Destinations do differ \emph{across categories} (\S\ref{sec:second}), but
that reflects how each category is sold, not a routing choice by the assistant.

The naive retail estimate is near zero for reasons of measurement, not behavior.
Matching the
placebo window removes a spurious negative; matching the URL path instead of the
host raises product-brand retail from $+0.11$ to $+0.52$~pp by catching the
product-page visits that carry the brand in the path; and brands that are
themselves retailers (Amazon, eBay, Target, Walmart) match a ``retail'' visit on
any visit to their own domain (much of it streaming or login), so we report them
separately and never treat them as the mentioned entity.

The retail figures answer different questions and should not be conflated. The
headline retail effect is the clean recommendation $\times$ non-customer lift,
$+1.0$~pp [$0.3$,$1.7$] (\S\ref{sec:clean}). The pooled all-mention lift is
$+0.52$~pp, of which about $+0.35$~pp is brand-specific net of the same-category
activity control (\S\ref{sec:catctrl}); it is confounded and is not the estimand we
report. The host-only and referrer-confirmed floors ($+0.11$~pp) are conservative
lower bounds on the same effect, not the clean $+1.0$.

\section{Boundary conditions}
\label{sec:second}

\paragraph{Generalization beyond the headline lexicon.} The headline analysis runs
on the cross-category lexicon (\S\ref{sec:data}). To test whether the effect is an
artifact of that brand set, we re-run the funnel on three \emph{fresh,
single-category} verticals (beauty, athletic apparel, audio/accessories) under the
same matched-placebo design (pooled over mentions, as in Table~\ref{tab:funnel}).
Table~\ref{tab:verticals} places those three alongside the cross-category lexicon
for reference. The recall step reproduces in every one, so the assistant-to-search
step is not specific to the headline lexicon; the downstream channel, by contrast,
is category-specific. The destinations differ: beauty routes to
retailers with a null own-site stage (it sells through specialty retail), apparel's
downstream is weak, and audio moves both. A single cross-category ``retail effect''
is the wrong target.

\begin{table}[t]
\caption{Generalization: the cross-category headline lexicon (top row, for
reference) versus three fresh single-category verticals. Within-user
backward-placebo funnel lift (pp, 7d, 95\% CI; pooled over mentions, as in
Table~\ref{tab:funnel}). Recall, the least-confounded stage (\S\ref{sec:naive}), is
significant in every row; the downstream channel is category-specific, so the
destination, not a single retail number, is the object of study.}
\label{tab:verticals}
\footnotesize
\setlength{\tabcolsep}{4pt}
\begin{tabular}{@{}lccc@{}}
\toprule
Lexicon & Recall & Discovery & Retail \\
\midrule
Cross-category    & $+2.08$ [$1.6$,$2.5$] & $+2.37$ [$1.7$,$2.9$] & $+0.52$ [$0.3$,$0.8$] \\
Beauty            & $+3.32$ [$1.6$,$5.8$] & $-0.29$ [$-1.9$,$0.8$] & $+2.92$ [$0.4$,$6.0$] \\
Apparel           & $+2.54$ [$0.9$,$4.3$] & $+1.67$ [$-0.1$,$3.3$] & $+0.06$ [$0.0$,$0.2$] \\
Audio/acc.        & $+1.45$ [$0.6$,$2.6$] & $+1.43$ [$0.5$,$2.6$] & $+0.77$ [$0.3$,$1.4$] \\
\bottomrule
\end{tabular}
\end{table}

\subsection{Assistant-level gaps are largely compositional}
\label{sec:assistant}

The clean acquisition effect is not evenly distributed across assistants. Split by
the assistant that named the brand, the recommendation $\times$ non-customer lift is
several times larger on ChatGPT than on Gemini at every stage
(Table~\ref{tab:assistant}); Claude is too sparse to estimate. The temptation is to
read this as ``ChatGPT recommendations work better.'' The within-user evidence points
the other way: the gap is mostly compositional, a difference in who uses each assistant.

Start with what it is not. Gemini's responses are not less commercial: they are if
anything more shopping-classified ($39.9\%$ vs $33.3\%$), at least as long, and name
brands at the same recommend-stance rate ($8.9\%$ each). So the content each surface
puts in front of a user is comparable. The behavior afterward is not. Restricting to
panelists who use \emph{both} assistants, the within-user ChatGPT-minus-Gemini gap is
not significant at any stage (Table~\ref{tab:assistant}, column~3): the same person,
handed the same kind of recommendation, shows no significant difference in response by
which assistant named it. The large between-user gap collapses within the user. What differs is who
sits in front of each surface.

\begin{figure}[t]
\centering
\includegraphics[width=\columnwidth]{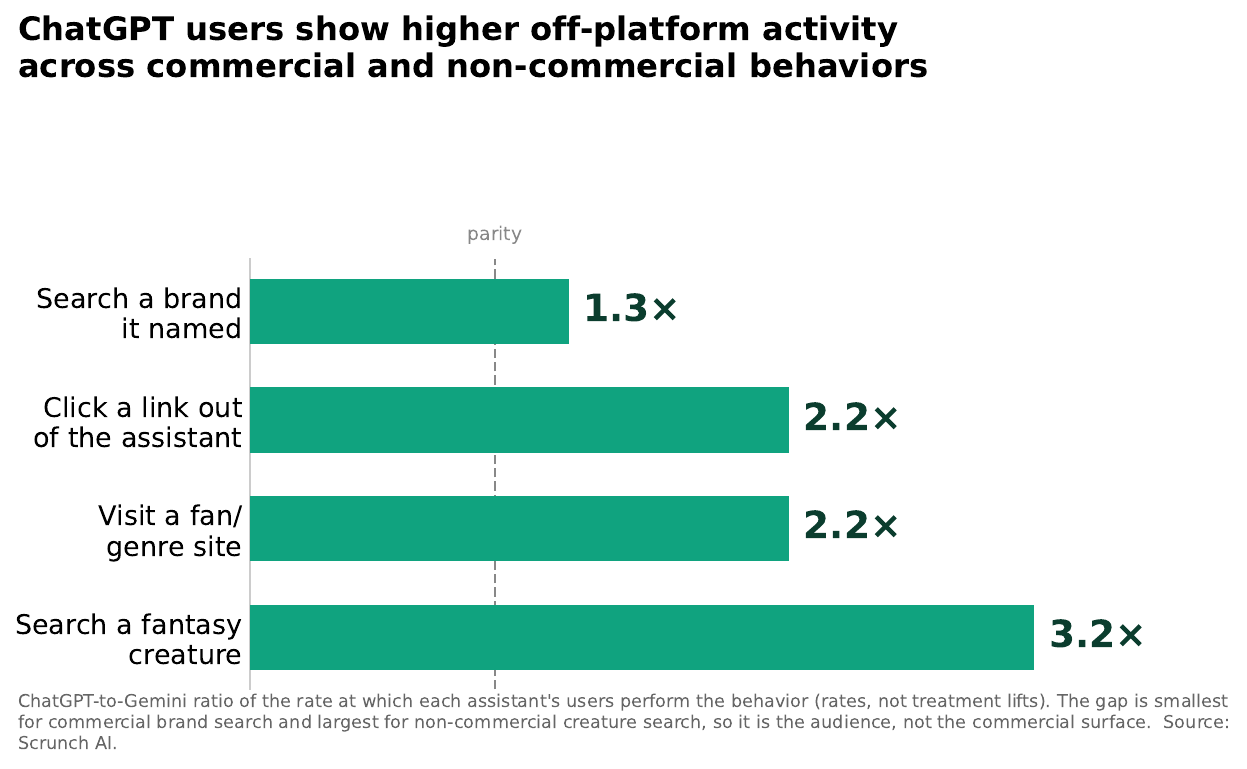}
\caption{Assistant-level gaps are largely compositional. ChatGPT-to-Gemini ratio of the
\emph{rate} at which each assistant's users perform four behaviors (rates, not
treatment lifts): searching a brand the assistant named, clicking any link out of
the assistant, visiting a fan or genre site, and searching a fantasy creature the
assistant mentioned (a non-commercial placebo, \S\ref{sec:assistant}). The gap is
smallest for commercial brand search ($1.3\times$) and largest for non-commercial
creature search ($3.2\times$), so it is a property of the audience, not the
commercial surface. These are between-user ratios; within-user, the click-out gap
does not persist (the same person clicks out no more on either assistant,
\S\ref{sec:assistant}).}
\Description{A horizontal bar chart of ChatGPT-to-Gemini rate ratios across four behaviors: searching a brand it named (1.3 times), clicking a link out of the assistant (2.2 times), visiting a fan or genre site (2.2 times), and searching a fantasy creature (3.2 times). All bars exceed parity at one, and the gap grows from the commercial brand search to the non-commercial creature search.}
\label{fig:assistant}
\end{figure}

This audience difference shows up well outside the commercial funnel
(Figure~\ref{fig:assistant}). ChatGPT users are about twice as likely to click any
link out of the assistant at all ($20.7\%$ of users vs $9.5\%$; $1.52$ vs $0.66$
outbound clicks per active user), so the same recommendation has a higher chance of
turning into a downstream visit simply because the person is more inclined to leave
the chat. This propensity is a trait of the person, not the interface. Among
panelists who use both assistants, a user's click-out rate on one predicts the rate
on the other (Spearman $+0.70$), and per response the same person clicks out no more
on ChatGPT than on Gemini. The between-user gap is a long tail of low-engagement
Gemini users, not a surface that elicits more clicks. A deliberately non-commercial
probe confirms the pattern is not about shopping. After an April 2026 reward-signal
episode made ChatGPT over-mention fantasy creatures across unrelated prompts,
independent of user demand~\cite{openai2026goblins}, we use those mentions as a
placebo lexicon with no commercial funnel. The same user-composition gap holds:
ChatGPT users search a named creature about $3.2\times$ as often as Gemini users and
visit fan or genre sites about $2.2\times$ as often (Figure~\ref{fig:assistant}).
The pattern is consistent with ChatGPT's users being more web-active overall, commercial or not, rather than with the assistant eliciting more action per response.

We therefore report the aggregate clean effect, pooled across assistants, as the
headline and treat the assistant split as an audience fact, and we make no
per-surface causal claim. In practical terms, the funnel's measurable effect
concentrates on ChatGPT today because of who uses it, not because its interface is
special. For measurement, the practical
implication is that assistant exposure should be evaluated jointly with the audience
composition of the surface where it occurs.

\begin{table}[t]
\caption{The ChatGPT--Gemini difference is consistent with audience composition.
Columns 1--2 split the clean recommendation $\times$ non-customer lift (pp, 7d,
95\% CI) by assistant: a large apparent gap. Column 3 restricts to panelists who use
\emph{both} assistants (a broader novelty panel, for power); the within-user gaps are
not significant at any stage, so we do not read the between-surface split as a causal
surface effect.}
\label{tab:assistant}
\small
\begin{tabular}{lccc}
\toprule
Stage & ChatGPT & Gemini & Within-user gap \\
\midrule
Recall    & $+6.6$ [$4.8$,$8.4$] & $+1.5$ [$0.3$,$2.7$]  & $-0.9$ [$-3.6$,$+1.9$] \\
Discovery & $+3.8$ [$2.3$,$5.7$] & $+0.9$ [$-0.2$,$1.9$] & $+2.2$ [$-0.5$,$+5.0$] \\
Retail    & $+1.7$ [$0.7$,$2.7$] & $+0.2$ [$-0.7$,$0.9$] & $+0.3$ [$-0.6$,$+1.2$] \\
\bottomrule
\end{tabular}
\end{table}

\paragraph{No familiarity gradient.} The long-tail case predicts larger effects
for less-familiar items. At the brand level, the correlation between an exogenous
familiarity measure (pre-period Wikipedia pageviews) and per-brand lift is near
zero at every stage (Spearman $+0.02$ to $+0.12$, none significant). We claim only
that, under this familiarity proxy and brand set, there is no reliable evidence the
per-mention lift is larger for less-familiar brands; if anything, moving behavior
mostly for salient brands would reinforce incumbents.

\paragraph{Presence, not position.} For the pooled recall lift, splitting by the
brand's order of first appearance in the response gives similar effects across rank
($+2.1/+2.0/+2.5$~pp for first/second/third-or-later mentions). Off-platform, in prose, being named at all, not
where in the list, is what moves behavior, unlike the position and popularity
biases documented for LLM-recommender ranked lists~\cite{lichtenberg2024}.

\paragraph{A spillover, and what it is not.} General retail activity rises after a
response even when the conversation is not about shopping ($+3.3$~pp at 24h,
rising over a week). This is \emph{not} a durable retail-specific shift: a neutral
control channel (news, social, video) rises at least as much over the same
horizons ($+8$~pp), so the broad multi-day elevation is generally heavier activity
after an assistant session, not induced commercial demand. What survives as
retail-specific is the shopping-versus-non-shopping differential ($+2.9$~pp at
24h, fading by a week), which a topic-blind activity burst cannot produce.

\section{Why the funnel runs through search}
\label{sec:mech}

The mechanism is consistent across results. The effect does not depend on where
the brand sits in the response, nor on the user clicking a link rendered inside
the answer. Direct click-through is negligible: in-answer links to brand or
retailer destinations account for only a tiny share of post-response brand
visits. The observed path is therefore best described as search-anchored rather
than assistant-click-driven. What matters is that the brand is
\emph{named}. The user reads a name they did not type, searches it, and lands on
a brand-identified destination, either the brand's own site or a retailer page
associated with that brand.

This is why last-click attribution misses the effect. By the time a query or
pageview is logged, the causal antecedent is two steps upstream and unrecorded.
The brand is reached through search
whether or not the landing URL itself carries the brand name.

\section{Deployment and implications for practice}
\label{sec:deploy}

\paragraph{Deployment.} The pipeline behind this paper is deployed production
measurement infrastructure at Scrunch~AI. It powers AI consumer-behavior tracking and
journey-intelligence analyses, largely in report form: how people move through
AI-mediated funnels, which prompts and personas appear at each stage, which brands
and sources are surfaced by assistants, and how AI exposure relates to downstream
open-web behavior. The version reported in this paper is not a standalone
customer-facing feature.

\paragraph{The production pipeline.} The research version reported here is a
disclosure-safe, aggregate view of that broader production measurement layer. It
joins LLM-conversation events to the same users' clickstream, extracts and
classifies brand mentions, separates product brands from platform and retailer
brands, applies path-aware URL matching, and computes matched-placebo behavioral
contrasts. These components feed downstream product metrics and customer-facing
analyses rather than appearing as a single UI control---the characteristic form of a
deployed measurement-infrastructure contribution~\cite{bakshy2014planout,tang2010overlapping}. The measurement has run in
production since early 2026, and this paper reports the disclosure-safe aggregate
estimates used in the identification design.

\paragraph{Implications for practice.} First, last-click tooling misattributes the
whole prompt-to-purchase path to organic search and ``direct'' retailer visits; an
honest accounting of an assistant's commercial role has to instrument the mention.
Second, the assistant's measurable effect is \emph{acquisition-like}: it concentrates
among users with no recent observed brand engagement, while among already-engaged
users it coincides with an active episode without clearly adding acceleration, so its
established role is reach. Third, a recommendation
moves two to three times more than an incidental name-drop, and the carrying channel
is category-specific, so measurement must track both own-site and retailer surfaces.
We report these as measurement implications, not as advice to manufacture mentions.
\ifvtwo
Translating these purchase-adjacent visits into lifetime value or acquisition cost is
the natural next step for practice. Pairing the reach we measure with downstream
conversion and retention data would let an advertiser price assistant-driven exposure
against existing channels, turning the incrementality estimate here into a bid-level
uplift signal; the transaction joins this requires are left to future work.
\fi

\section{Discussion and Limitations}
\label{sec:limits}

Conversational assistants create unlogged brand exposures that move
observably-unengaged users into open-web brand navigation. The path from prompt to
commerce runs through the open web, and most of it is
unlogged. Once the existing-customer episodes and incidental name-drops are
separated out, the measurable pattern is \emph{acquisition-like reach}: the
assistant moves observably-unengaged users to search, to the brand's site, and to a
retailer, and last-click analytics attributes almost none of it to the assistant.

\paragraph{What we do not claim.} The panel observes web navigation, not attention: a
user may read a recommendation and never search, search in a channel we do not see,
act inside an app, buy offline, or move outside the seven-day window, so the rates are
a lower bound on influence. ``Non-customer'' means no recent \emph{observed}
engagement, not the absence of any brand relationship. We do not observe transactions; retail is a
purchase-adjacent product-page visit, and ``purchase'' refers to that funnel
direction, not a logged sale. The design is observational. Our defenses against
confounding are layered: the same-category control (category intent and activity
surges), the pre-trend event-study and non-customer conditioning
(existing-customer behavior), the stance split, and the no-prior-same-category
stratum (\S\ref{sec:catctrl}). None fully rules out a \emph{brand-specific} intent
shock that, in the same session, both prompts a recommendation and drives action.
The same-category stratum narrows it sharply, since the effect holds for users with
no observed prior same-category lexicon-brand engagement, yet a within-session shock specific to the
named brand remains unidentifiable on observational data; a randomized or
encouragement design would be needed to close it. Among existing customers the
mention coincides with an active episode and adds little beyond it in the pooled case
(3-day increment $\approx +0.4$~pp discovery), but the recommend-to-existing subset
is too small to resolve a modest acceleration. The familiarity proxy is encyclopedic,
not commercial. On
multiplicity, the three stages and the recommendation$\times$non-customer contrast
are pre-specified, and we report every cell of the decomposition rather than
selecting among them; intervals are not multiplicity-adjusted, which the nested
structure of the stages makes defensible but worth flagging. Brand-clustered and
two-way (user-or-brand) bootstraps leave the headline intervals clear of zero, so
the result is not an artifact of clustering only on the user (appendix). Stance is assigned by
a small language model on a brand-local snippet; an independent re-classification
of a held-out mention sample with a stronger model agrees on $90\%$ (Cohen's
$\kappa=0.62$), with disagreements at the recommend/neutral boundary and
essentially none between recommend and caution, so classifier error likely
attenuates the recommend/neutral contrast, which makes the
reported recommend dose conservative. Finally, the headline lexicon is high-recognition
consumer brands in two English-speaking markets; the recall replication across three fresh
single-category verticals and the absence of a
familiarity gradient speak to generalization, but the long tail of rarely-mentioned,
niche brands remains out of reach here. Per-assistant lifts differ in raw form but attenuate to
non-significance within users, so we treat the assistant comparison as confounded
by self-selection into surfaces and report the aggregate clean effect (pooled across
assistants) rather than a per-surface causal estimate (\S\ref{sec:second}).

\ifshowethics
\section*{Ethics and Human-Subjects Statement}
This is secondary analysis of previously collected, opt-in, de-identified
web-browsing, search, and conversation events; panelists consented to behavioral
measurement under the panel provider's terms, and the authors did not intervene on
user experience or contact participants. No new data were collected for this study;
the analysis uses previously collected, de-identified, opt-in panel data with no
intervention on or contact with participants, and the authors treated the work as
secondary aggregate measurement under the panel provider's consent and governance
process. Results are reported only in aggregate;
no raw conversations, searches, individual URLs, or sample-size statistics that
could expose panel composition are released. One contextual point deserves naming:
joining conversation content to browsing and search behavior is more sensitive
than either stream alone, and our use is confined to aggregate measurement. The
normative implication is that assistants are becoming commercial gatekeepers whose
entity choices allocate attention and, as we show, can move previously-unengaged
users toward the brands they name; we frame this as a case for measurement and
governance of AI-mediated brand exposure, not as advice to optimize for mentions.

\begin{acks}
AI coding assistants helped write and refactor the analysis and figure-generation
scripts. The study design, identification strategy, classifier choices, and every
reported estimate were specified and verified by the authors, who take full
responsibility for all claims, numbers, and citations.
\end{acks}
\fi

\bibliographystyle{ACM-Reference-Format}
\bibliography{references}

\appendix
\section{Secondary and diagnostic results}

\begin{table}[!ht]
\caption{Retail measurement failure modes and corrections. Host-only matching reads
near-null; the corrections recover a real but modest brand-specific retail effect.}
\label{tab:retailfail}
\small
\begin{tabular}{p{0.22\linewidth}p{0.30\linewidth}p{0.34\linewidth}}
\toprule
\textbf{Failure mode} & \textbf{Example} & \textbf{Correction} \\
\midrule
Host-only matching & a bare \texttt{amazon.com} visit & match the brand in the URL \emph{path} \\
Retailer-as-brand & Amazon, eBay, Target, Walmart & report separately; never treat as the mentioned entity \\
Parent-brand domain & \texttt{iphone}$\rightarrow$\texttt{apple.com} & parent-domain mapping ($+5.21$~pp for those tokens) \\
Homograph string & \texttt{canon}, \texttt{stanley} & curated lexicon and homograph pruning \\
\bottomrule
\end{tabular}
\end{table}

Stance is classified by a small language model (recommend / neutral / caution) on
a brand-local snippet; most mentions are neutral, recommendations a minority (under
one in ten), and caution rare. Representative stances are a recommendation
(``consider the Garmin Forerunner''), a neutral reference (``you can use Samsung
Pay,'' ``your Netflix download''), and a caution (a warning or unfavorable
comparison). Audited against an independent stronger-model re-classification
(Table~\ref{tab:stance}; agreement $90\%$, $\kappa=0.62$, model-vs-model not human
ground truth), the off-diagonal mass
falls at the recommend/neutral boundary and recommend is almost never confused with
caution, so classifier error likely attenuates the recommend/neutral contrast
rather than inflating it. The classifier is a noisy measurement layer, good enough
to separate the poles but not a ground truth.

\begin{table}[!ht]
\caption{Stance audit (model-vs-model agreement, not human ground truth):
small-model stance (columns) against an independent
stronger-model re-classification (rows) on a held-out mention sample, as
row-normalized percentages (each row sums to $100\%$). Agreement
$90\%$, Cohen's $\kappa=0.62$. Off-diagonal mass concentrates at the
recommend/neutral boundary; recommend and caution are almost never confused, so
error pushes the recommend dose toward neutral, making the reported effect conservative.}
\label{tab:stance}
\small
\begin{tabular}{lccc}
\toprule
 & \multicolumn{3}{c}{small model (row \%)} \\
strong model & recommend & neutral & caution \\
\midrule
recommend & $75.1$ & $24.9$ & $0.0$ \\
neutral & $5.3$ & $93.4$ & $1.2$ \\
caution & $4.1$ & $36.9$ & $59.0$ \\
\bottomrule
\end{tabular}
\end{table}

The clean recommendation$\times$non-customer cell is stable along three further
axes. Clustering the bootstrap on brand rather than user, or two-way on user and
brand, leaves the point estimates unchanged and the intervals clear of zero
(brand-clustered recall $[+3.0,+5.9]$, discovery $[+1.1,+3.8]$, retail
$[+0.2,+1.9]$); collapsing to the first mention per user-brand gives
$+4.7/+2.6/+1.1$. Leave-one-category-out, dropping each substitute set in turn,
holds recall in $[+3.6,+4.8]$, discovery in $[+2.1,+2.8]$, and retail in
$[+0.8,+1.1]$, so no single category carries the effect. Swapping the placebo
offsets ($\{14,21,28\}$d by default) for $\{21,28,35\}$, $\{14,21\}$, or
$\{7,14,21\}$ holds recall in $[+4.0,+5.2]$ and discovery in $[+2.3,+3.4]$, with
retail positive and significant except under the two-offset set, where it grazes
zero. The funnel also survives dropping homograph-prone strings, user- and
brand-normalized reweighting, and leave-one-brand-out. Brand matching is a
substring match; restricting the recall outcome to word-boundary search matches
leaves the pooled all-mention recall estimate unchanged ($+2.08$~pp), and the few substring-prone strings
(\texttt{kia}$\subset$\texttt{nokia}, \texttt{uber}$\subset$\texttt{tuberculosis},
\texttt{ipad}$\subset$\texttt{tripadvisor}) only attenuate it---dropping them raises
recall to $+2.49$~pp---because the matched placebo nets out baseline homograph
searches and mention-side substring noise adds only null rows. For product tokens
whose owned site is a
parent-brand domain (\texttt{iphone}$\rightarrow$\texttt{apple.com}), substring
own-domain matching undercounts discovery; parent-domain mapping yields
$+5.21$~pp [$3.95$, $6.59$] there, so the substring headline is conservative. The search-anchored retail measure---which credits a retailer visit to the mention
when a same-name search shortly precedes it (\S\ref{sec:measure})---fails a
direction test: the forward anchor lifts retail $+0.28$~pp, but a reverse-order
placebo lifts it a comparable $+0.36$~pp and a shifted anchor is null, so the
association is not specific to the causal order. The headline therefore excludes
this search-anchored component and rests on path-aware retail
only. A referrer-confirmed retailer variant gives a directional lower bound
($+0.11$~pp [$0.01$, $0.22$]).

\end{document}